\documentclass[aps,superscriptaddress,reprint]{revtex4-2}

\usepackage[utf8]{inputenc}
\usepackage{xcolor}
\usepackage{pst-node}
\usepackage{soul}
\usepackage{graphicx}
\usepackage{amsmath, amssymb, amsthm}
\usepackage{mathtools}
\usepackage{hyperref}
\usepackage[shortlabels]{enumitem}
\usepackage{dsfont}
\usepackage{amsthm}
\usepackage{amssymb}
\usepackage{amstext}
\usepackage{braket}
\usepackage{amsfonts}
\usepackage{nicefrac}
\usepackage{extarrows} 
\usepackage{graphicx}
\usepackage{relsize}
\usepackage{soul}
\usepackage{xfrac} 

\newcommand{\skiptext}[1]{}

\newcommand{\Fig}[1]{Fig.~\ref{#1}}

\begin{document}

\title{Parallel implementation of the Density Matrix Renormalization Group method 
achieving a quarter petaFLOPS performance on a single DGX-H100 GPU node}
 
\author{Andor Menczer}
\affiliation{%
Strongly Correlated Systems Lend\"ulet Research Group,
Wigner Research Centre for Physics, H-1525, Budapest, Hungary
}%
\affiliation{%
Eötvös Loránd University, Budapest, Hungary
}%

\author{Maarten van Damme}
\affiliation{SandboxAQ, Palo Alto, California, USA}
\author{Alan Rask}
\affiliation{SandboxAQ, Palo Alto, California, USA}
\author{Lee Huntington}
\affiliation{SandboxAQ, Palo Alto, California, USA}

\author{Jeff Hammond}
\email{jeffpapers@nvidia.com}
\affiliation{%
NVIDIA Helsinki Oy, Porkkalankatu 1, 00180 Helsinki
}%

\author{Sotiris S. Xantheas}
\email{Sotiris.Xantheas@pnnl.gov}
\affiliation{%
Advanced Computing, Mathematics, and Data Division, Pacific Northwest National Laboratory, Richland, Washington 99354, USA}%
\affiliation{%
Department of Chemistry, University of Washington, Seattle, WA 98195, USA}

\author{Martin Ganahl}
\email{martin.ganahl@sandboxaq.com}
\affiliation{SandboxAQ, Palo Alto, California, USA}%

\author{\"Ors Legeza}
\email{legeza.ors@wigner.hu}
\affiliation{%
Strongly Correlated Systems Lend\"ulet Research Group,
Wigner Research Centre for Physics, H-1525, Budapest, Hungary
}%
\affiliation{Dynaflex LTD, Zrínyi u 7, 1028 Budapest, Hungary}
\affiliation{
Institute for Advanced Study,Technical University of Munich, Germany, Lichtenbergstrasse 2a, 85748 Garching, Germany
}
\affiliation{Parmenides Stiftung, Hindenburgstr. 15, 82343, Pöcking, Germany}

\date{\today}

\begin{abstract} 
We report cutting edge performance results for a hybrid CPU-multi GPU implementation of the spin adapted \textit{ab initio} Density Matrix Renormalization Group (DMRG) method
on current state-of-the-art NVIDIA DGX-H100 architectures. We evaluate the performance of the DMRG electronic structure calculations for the active compounds of the FeMoco and cytochrome P450 (CYP) enzymes with complete active space (CAS) sizes of up to 113 electrons in 76 orbitals [CAS(113, 76)] and 63 electrons in 58 orbitals [CAS(63, 58)], respectively. We achieve 246 teraFLOPS of sustained performance, an improvement of more than 2.5$\times$ compared to the performance achieved on the DGX-A100 architectures and an 80$\times$ acceleration compared to an OpenMP parallelized implementation on a 128-core CPU architecture. Our work highlights the ability of tensor network algorithms to efficiently utilize high-performance GPU hardware and shows that the combination of tensor networks with modern large-scale GPU accelerators can pave the way towards solving some of the most challenging problems in quantum chemistry and beyond.
\end{abstract}

\maketitle

\emph{Introduction} 

Our current understanding of the properties of materials and molecules rests on the foundations of quantum mechanics. Many modern technologies -- such as semiconductor devices \cite{bardeen_1948}, magnetic resonance imaging (MRI), nuclear power or photovoltaic cells -- would be impossible without the fundamental understanding of the underlying quantum mechanical effects governing the processes that are responsible for the development of these technologies. 
The properties of any
molecule or material can in theory be computed from solutions 
of the Schr\"odinger equation, but obtaining the exact solution is in general impossible except in rare special cases \cite{francesco2012conformal}, leaving scientists with the need to settle for approximations. 
The exponential growth in computational power over the last few decades has led to approximate numerical methods 
that have become the predominant choice for modeling materials and molecules in both scientific and industrial applications. Prominent examples include density functional theory (DFT) \cite{kohn_1965, Weitao_2012, Becke_2014, jones_2015} (which has become a standard tool in the scientific community and beyond 
\cite{neese_orca_2020,nwchem,molpro,Gaussian,qchem, ADF2001, VASP, pyscf, terachem, quantum_espresso_1, quantum_espresso_2, quantum_espresso_3, pederson_large}), single and multi-reference Coupled Cluster (CC)\cite{Cizek_CC, Bartlett_CC,  Bartlett-2007, Shavitt_Bartlett_2009, Bartlett_2024, Lyakh_MRCC_review, evangelista_perspective_2018} approaches, quantum Monte Carlo (QMC) \cite{booth_fermion_2009, FCIQMC2, FCIQMC3, FCIQMC4, AFQMC, AFQMC2, AFQMC_jctc} and various other approximations of
full configuration interaction (FCI)\cite{CIPSI1, CIPSI2, HBCI, HBCI2, iCI, iCI2, Zimmerman-2017a, Zimmerman-2017b, MBE_CI, SHCI, ACI, ASCI}, or tensor networks \cite{affleck_rigorous_1987, fannes_finitely_1992, White-1992b, White-1993, nishino_density_1995, ostlund_thermodynamic_1995, rommer_class_1997, White-1999, Schollwock-2005, Legeza-2008, CHAN2009, Schollwock-2011, chan_review, Szalay-2015a,Orus-2019,Chan-2020,Baiardi-2020}, 
to name a few. 

Despite the tremendous algorithmic and hardware advances over the last half century, many quantum mechanical phenomena in chemistry, material science, and condensed matter physics are still not thoroughly understood. Examples include the mechanism of action of biological enzymes \cite{reiher_elucidating,li_electronic_2019, Goings-2022, weser_chemical, sono_heme, Schoneboom_elusive, meunier_mechanism,phung_toward,li_manni_understanding,li_manni_combining, li_manni_role}, the properties of exotic phases of matter~\cite{topo_review, yan_spin_liquid, kitaev_2001} (including the debated existence of anyonic quasi-particles \cite{Alicea_2012, KITAEV20032}), or even the exact mechanisms of observed cases of high-temperature superconductivity in certain materials \cite{supercond_review, bednorz_possible_1986, twisted_bilayer_graphen}.
A common theme among these phenomena is that they all require the solutions of the many-body Schr\"odinger equation to obtain a proper understanding of their electronic structure in their ground and excited electronic states. In this context, tensor networks have emerged as one of the most powerful numerical approaches for tackling these challenging problems \cite{White-1993, chan_highly_2002,Legeza-2003a,yan_spin_liquid, vidal_mera, verstraete2004renormalizationalgorithmsquantummanybody}. Tensor networks are a class of many-body wave functions that can be efficiently stored and manipulated using classical hardware. Tensor networks can parameterise wave functions obeying an area law of entanglement \cite{Eisert-2010} with possibly logarithmic corrections \cite{vidal_entanglement, vidal_mera}, and can be combined with local unitary optimization to reduce entanglement \cite{Krumnow-2016, friesecke2024globalfermionicmodeoptimization, Menczer-2024}. They are also ideally suited to parameterize ground states of gapped, local quantum systems in 1d and 2d. The most successful tensor network, the matrix product state (MPS), is arguably the gold standard approach for obtaining ground states of strongly correlated quantum systems in 1d and 2d \cite{yan_spin_liquid}. In the area of quantum chemistry, the density matrix renormalization group (DMRG) algorithm, a variational optimization algorithm over the space of MPS, has emerged at the forefront of strongly correlated electron methods, and is widely regarded as a gold standard method for systems encompassing multi-reference character
\cite{White-1999, Legeza-2008, Krumnow-2016, Szalay-2015a, CHAN2009, chan_review, gunst-2018, Gunst-2019, Baiardi-2020, Cheng-2022}.

The core operations required in the vast majority of all tensor network algorithms are tensor contraction and matrix factorization, both of which are highly amenable to parallelization and Graphics Processing Unit (GPU) acceleration \cite{Hager-2004,Stoudenmire-2013,Nemes-2014,Ganahl-2019, Milsted-2019,Brabec-2021,Zhai-2021,Gray-2021,Unfried-2023,Ganahl-2023,Menczer-2023a,Menczer-2023b,Menczer-2024,Xiang-2024}. 
In this context, growing attention is being focused towards developing novel tensor network algorithms that can efficiently utilize highly specialized Artificial Intelligence (AI) accelerators. Examples include recent work on 
$SU(2)$ spin adapted implementations of DMRG run on NVIDIA DGX-A100~\cite{nvidia-a100} architectures~\cite{Menczer-2023a,Menczer-2023b, Menczer-2024}, or multi-node multi-GPU architectures~\cite{Menczer-2023d,Xiang-2024}. 
\newline
In this work we report on recent progress using large-scale GPU hardware to substantially accelerate tensor network simulations for quantum chemistry and materials science applications. Benchmark calculations of our highly-parallelized, GPU-accelerated and SU(2)-aware implementation of the DMRG algorithm on NVIDIA DGX-H100 GPU supercomputers have achieved sustained performance of $\sim$250 teraFLOPS (trillion floating-point calculations per second) which represents an 80$\times$ speedup compared to a state-of-the-art implementation on a traditional Central Processing Unit (CPU) executed on a 128-core CPU architecture.
\\

\emph{Numerical procedure} 

In the following we will discuss performance benchmarks for the
DMRG method for quantum chemistry applications. The DMRG algorithm is the oldest and most important tensor network algorithm, and can be understood as a variational method in the space of so-called matrix product state (MPS) \cite{Verstraete-2023} wave functions. In  
the quantum chemistry context, an MPS is a parameterization of a many-body wave function in terms of $N$ spinful orbitals $\ket{i_n}$ using $N$ order-3 tensors $A^{i_n}_{\alpha_{n-1}\alpha_n}$ of dimension $(D_{n-1},4,D_n)$, i.e.
\begin{equation}
  \ket{\Psi_{MPS}}  = \sum_{\{i_k\}} \sum_{\{\alpha_p\}}[A_1]_{1\alpha_1}^{i_1} [A_2]_{\alpha_1\alpha_2}^{i_2} \dots [A_{N}]_{\alpha_{N-1}1}^{i_{N}} \ket{i_1\dots i_k}
\end{equation}
where the first and the last are order-2 tensors or matrices. 
The DMRG algorithm can be used to construct a variational approximation to the ground state of the quantum chemistry Hamiltonian $H$ over the space of MPS, i.e. 
\begin{equation}
  E_{opt} = \min_{\ket{\Psi_{MPS}}}\frac{\bra{\Psi_{MPS}}H\ket{\Psi_{MPS}}}{\braket{\Psi_{MPS}|\Psi_{MPS}}}.
\end{equation}
The bond dimension $D\equiv\max(\{D_n\})$ controls the accuracy of the approximation (larger $D$ is better), with values of $D\sim\mathcal{O}(10^4)$ often mandatory to reach sufficient accuracy in quantum chemistry applications. The computational complexity and memory requirements of DMRG scale as $D^3N^4$  and $D^2N^2$, respectively. The DMRG algorithm performs an iterative optimization (one MPS tensor update at a time) of the wave function, where each update is obtained by solving a large hermitian eigenvalue problem using e.g. the Lanczos or Davidson method. This step usually accounts for 80\% of the execution time, and scales as $\mathcal{O}(D^3)$.
One sequence of updates of all tensors is called a DMRG sweep.
For more details on the DMRG and tensor networks in general, we refer the reader to the existing literature \cite{Schollwock-2005, Schollwock-2011,Noack-2005,Szalay-2015a,Chan-2008,Orus-2014,Baiardi-2020, verstraete_matrix_2008, RevModPhys.93.045003}.
\\

\emph{Performance assessment}

In the following we present performance benchmarks of DMRG-CAS($M,N$) of $M$ electrons in $N$ active orbitals on DGX-H100 \cite{nvidia-dgx-h100} for a series of increasingly complex molecular systems,  namely F$_2$ [CAS(18, 18)]~\cite{Legeza-2003a}, N$_2$ [CAS(14, 28)]~\cite{Chan-2004b}, the Iron-Molybdenium cofactor [FeMoco, CAS(54, 54)~\cite{Reiher-2017} and CAS(113, 76)~\cite{Li-2019}], and 
the activated heme group of cytochrome P450 [CAS(63, 58)~\cite{Goings-2022}]. All bond dimensions $D$ are reported as SU(2) multiplets, with the corresponding U(1) bond dimensions indicated separately where applicable.

\begin{figure}
    \centering    
    \includegraphics[width=0.48\textwidth]{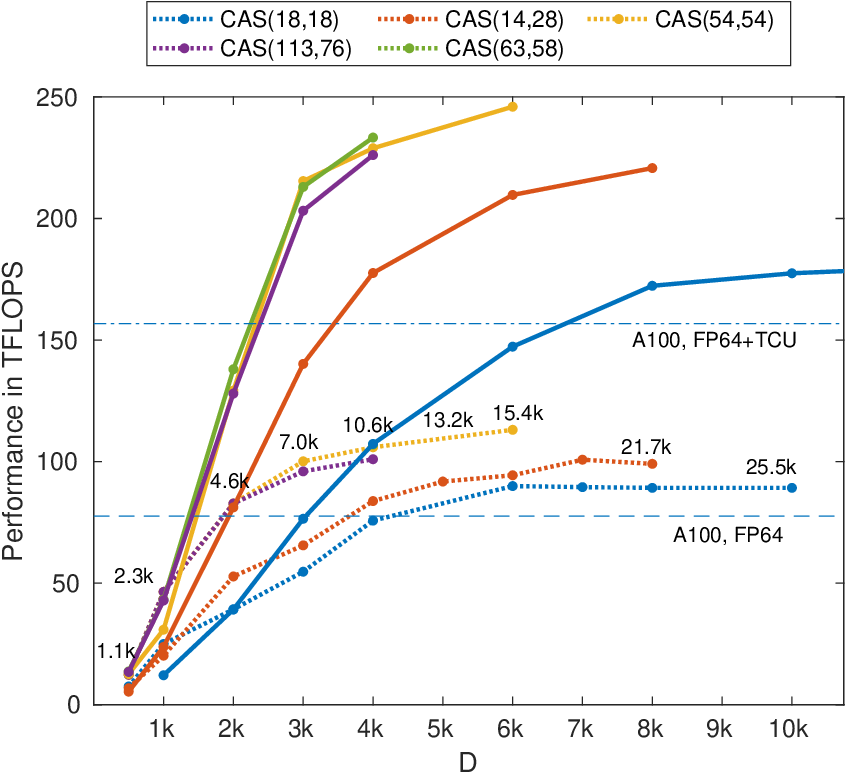}
    \caption{Benchmark results obtained via the SU(2) spin-adapted hybrid CPU plus multi-GPU DMRG calculations for the F$_2$ molecule on a CAS(18,18) orbital space~\cite{Legeza-2003a}, the
    N$_2$ molecule on a CAS(14,28) space~\cite{Chan-2004b}, FeMoco on CAS(54,54)~\cite{Reiher-2017} and CAS(113,76)~\cite{Li-2019} spaces, and P450 on CAS(63,58)~\cite{Goings-2022}.
    The solid lines correspond to calculations performed on a DGX-H100 system.
    As a reference, the dotted lines trace the results obtained on a DGX-A100 system.
    The estimated FP64 theoretical upper bound for DGX-A100 
    is shown by the horizontal dashed line, while the same but also including specialized tensor core units (TCUs) by the horizontal dashed-dotted line. 
    Numbers indicate the corresponding $U(1)$ bond dimension values.
    }
    \label{fig:perf}
\end{figure}

In \Fig{fig:perf} we show the performance results of our DMRG implementation on the above mentioned systems, and for increasing values of SU(2) bond dimension $D$. For all simulations we observe an initial linear increase in performance with 
increasing $D$ and a problem-dependent saturation value. For the smallest systems [CAS(18,18)], the performance saturates at $\sim$180 teraFLOPS. For the largest systems [CAS(54,54) and CAS(63, 58)] we achieve sustained performance of $\sim$250 teraFLOPS and expect to reach the performance plateau between $D\approx8000$-$10000$. Beyond these bond dimensions, host-device data-communication starts to become the dominating factor due to memory limitations on the DGX-H100 and causes a performance breakdown for these large CAS DMRG simulations. However, we expect that MPI-based approaches~\cite{Xiang-2024,Menczer-2023c} and advanced hardware (such as GH200~\cite{gh200} or AMD MI300~\cite{mi300} superchips) will mitigate this problem and allow us to scale simulations well into and eventually surpassing this regime.
Indeed, for GH200 and MI300 hardware, the CPU and GPU have direct shared-memory access across the node, largely eliminating the host-device communication bottleneck. For a more detailed discussion on the nature of the CAS-size dependence of the performance plateau values, we refer the reader to Reference~\cite{Menczer-2023a}. 

In summary, we observe an almost ideal 2.5$\times$ increase in performance compared to DGX-A100 (dashed lines in \Fig{fig:perf}) and an 80$\times$ increase compared to other state-of-the-art OpenMP parallelized implementations of quantum-chemistry DMRG calculations on 128 CPU cores~\cite{Menczer-2023a}. Two key hardware features that allow us to achieve such performance gains are the massive compute throughput and high memory bandwidth on DGX-H100, as well as the availability of efficient implementations of core linear algebra subroutines in NVIDIA math libraries (CUBLAS).

\begin{figure}
    \centering    
    \includegraphics[width=0.48\textwidth]{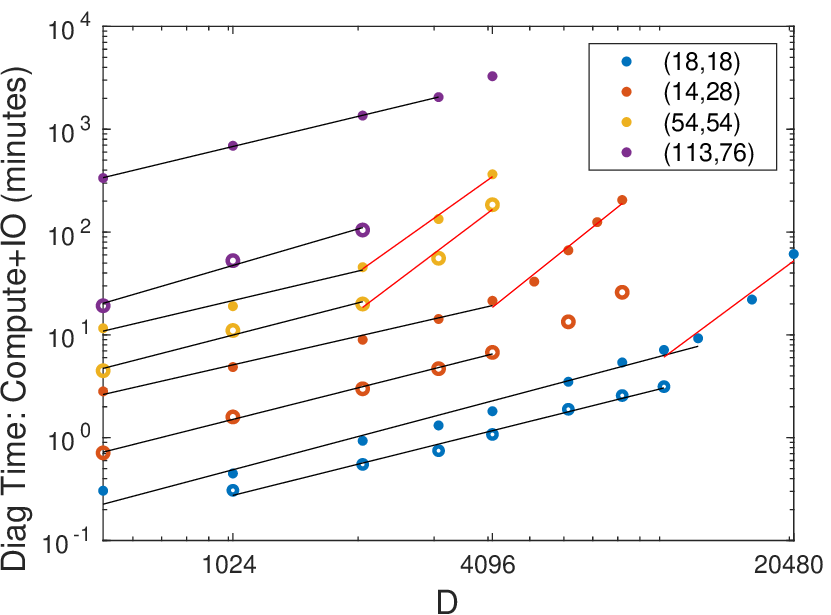}
    \caption{Total diagonalization time of seven DMRG sweeps for the eight GPU accelerated diagonalization procedure measured in minutes including host-device IO overhead for the F$_2$ CAS(18,18), N$_2$ CAS(14,28), FeMoco CAS(54,54) and CAS(113,76) as a function of DMRG bond dimension on A100 (solid dot symbol, $\bullet$) and on H100 (open symbol, $\circ$) architectures. 
    The solid lines are results of ﬁrst-order polynomial ﬁts on selected data sets corresponding to measured performance up to saturation of GPU performance (black) and for region where performance saturated (red). The fitted exponents for the H100 calculations are $1.05\pm 0.1$ and $2.95\pm 0.2$, respectively.  
    }
    \label{fig:time}
\end{figure}

In \Fig{fig:time} we show the
total wall time spent in the Davidson diagonalization (including host-device communication) over seven DMRG sweeps as a function of bond dimension $D$ for the systems considered above. Consistent with \Fig{fig:perf}, we observe a linear increase in the wall time for bond dimensions below the performance plateau, and the expected $D^3$ scaling once we reach the performance plateau. 

Due to the high performance of the latest generations of GPUs and the high degree of parallelization of our DMRG implementation, the wall time for the diagonalization step is reduced to a point where it is no longer the limiting operation, and instead data transfer operations between CPU memory and storage media become the bottleneck. We utilize data compression techniques and asynchronous data transfer approaches to partially mitigate this problem, at the cost of increasing memory requirements by $\sim$30\%. For multi-node systems, distributed data approaches \cite{Ganahl-2023,Brabec-2021,Xiang-2024,Menczer-2023c}
can be used to mitigate similar data transfer bottlenecks.
\\

\emph{Spin states of Cytochrome P450 heme group} 

In the following we present DMRG-CAS results for the low-lying spin states of the heme-group of the Cytochrome P450 (3A4 isoform) enzyme in its active state (Cpd I) \cite{Goings-2022}. Cytochromes are heme-containing enzymes primarily responsible for detoxification of organisms \cite{Goings-2022, Lynch-2007,Zhao-2021}, where the heme-group, an iron porphyrin system, is responsible for catalyzing chemical reactions with substrates of the enzyme. Iron porphyrin structures appear as key building blocks in various enzymes. In the active state of P450 (3A4), the iron-porphyrin ring has an oxygen and cysteine bound to the central Fe atom above and below the iron-coordinating plane. Their low-lying energy spectrum features three nearly degenerate states with spin $s=\sfrac{1}{2}$, $\sfrac{3}{2}$ and $\sfrac{5}{2}$, whose relative energies
depend on the geometry and the local chemical environment of the heme group. A full understanding of the electronic structure of this system remains an open problem \cite{weser_chemical, sono_heme, Schoneboom_elusive, meunier_mechanism,phung_toward,li_manni_understanding,li_manni_combining, li_manni_role, rask2022many}. The multi-reference character and the near-degeneracy of the doublet ($s=\sfrac{1}{2}$) and quartet ($s=\sfrac{3}{2}$) states \cite{weser_chemical} pose significant challenges for existing computational approaches, with large active spaces being crucial for obtaining qualitatively and quantitatively accurate results \cite{weser_chemical, li_manni_understanding}.
Here, we revisit the problem of computing the DMRG energies of the $s=\sfrac{1}{2}$, $\sfrac{3}{2}$, and $\sfrac{5}{2}$ states for the active spaces defined in \cite{Goings-2022}, and extend them to the CAS(63, 58) space. To the best of our knowledge, this is the largest DMRG-CAS calculation reported to date for this compound. 
Our primary aim is to demonstrate the ability of our our SU(2) symmetric, GPU-accelerated DMRG implementation to run calculations on very large active spaces with large bond dimension within a significantly reduced runtime on DGX-H100 machines.

\begin{figure}
    \centering    
\includegraphics[width=0.48\textwidth]{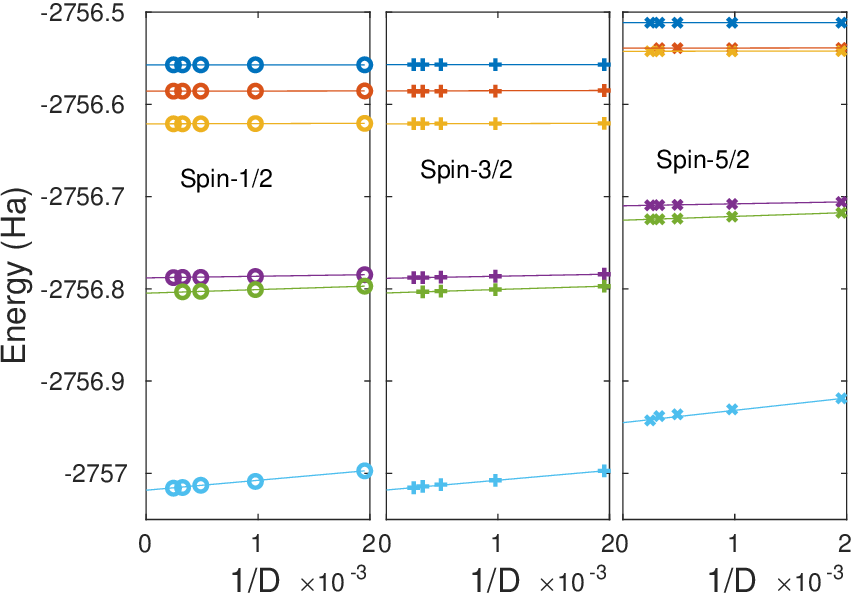}
    \caption{Scaling of the energy for spin states with total spin $\sfrac{1}{2}$ (left panel), 
    $\sfrac{3}{2}$ (middle panel) and 
    $\sfrac{5}{2}$ (right panel) as a function of the inverse DMRG SU(2) bond dimension for the Cytochrome P450 enzyme for the model spaces of 
    CAS(17,15),
    CAS(25,23),
    CAS(33,31),
    CAS(45,41),
    CAS(47,43), and
    CAS(63,58) introduced in ~\cite{Goings-2022}.
    Solid lines are the result of a second order polynomial fits.
    }
    \label{fig:p450-scaling}
\end{figure}

\begin{figure}
    \centering    
\includegraphics[width=0.48\textwidth]{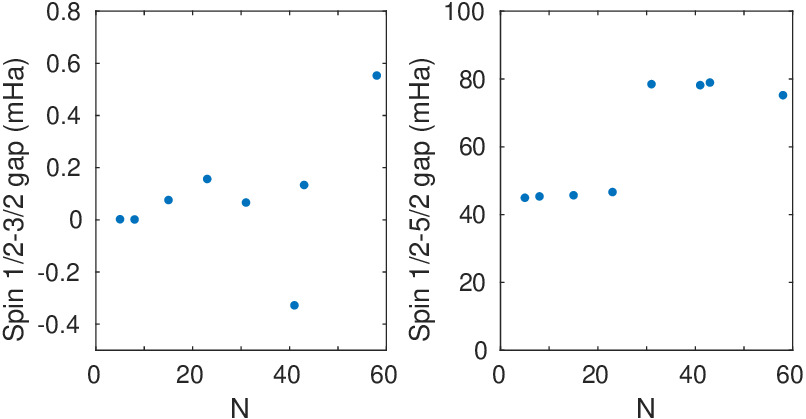}
    \caption{The $D\rightarrow\infty$ extrapolated spin gap (mHartree) measured between the spin $\sfrac{1}{2}$ ground and spin $\sfrac{3}{2}$ excited states (left panel) and between the spin $\sfrac{1}{2}$ ground and spin $\sfrac{5}{2}$ excited states (right panel) as a function of model CAS spaces with increasing complexity, i.e. with increasing number of orbitals and number of electrons~\cite{Goings-2022}.
    }
    \label{fig:p450-gap}
\end{figure}

In \Fig{fig:p450-scaling} we present  
the $1/D$ scaling analysis of the calculated energies for the lowest lying eigenstates with total spin $\sfrac{1}{2}$ (left panel), 
$\sfrac{3}{2}$ (middle panel), and 
$\sfrac{5}{2}$ (right panel) used to obtain the truncation free extrapolated $D\rightarrow \infty$ limit~\cite{Schollwock-2005} for the different CAS spaces (solid lines are second order polynomial fits). 
The extrapolated energies for the spin $\sfrac{1}{2}-\sfrac{3}{2}$ and $\sfrac{1}{2}-\sfrac{5}{2}$ gaps are shown for increasing CAS space sizes in Fig.~\ref{fig:p450-gap}. We observe a degeneracy on the order of 0.1 mHartree for the doublet and quartet states, which lies within the established accuracy of the largest measured DMRG truncation error (order $10^{-2}$ mHartree for $D\leq4096$, i.e. for $D_{U(1)}~\simeq 11000$).
The spin $\sfrac{1}{2}-\sfrac{5}{2}$ gap (right panel of \Fig{fig:p450-gap}) remains positive and the spin $\sfrac{5}{2}$ state lies above both the spin $\sfrac{1}{2}$ and spin $\sfrac{3}{2}$ states. 
However, in order to resolve the spin gaps at a high level of accuracy (or even capture the qualitative behaviour) it is imperative to extend the calculations to larger active spaces while including dynamical correlation effects (e.g. via NEVPT2, the tailored coupled-cluster (TCC)\cite{Veis-2016} or the restricted active space DMRG-RAS-X~\cite{Barcza-2022ras,Friesecke-2023} methods). Namely, the resolution of the spin gaps is highly dependent upon a balanced treatment of static and dynamic correlation effects for all three spin states. We have also performed DMRG-CASSCF \cite{Zgid-2008c} calculations on this system, using smaller active spaces, which yield substantially lower energies compared to calculations with fixed non-optimized orbitals. This is part of currently ongoing research and will be published in the near future.

We emphasise again that the high performance of our DMRG implementation, in particular observed already at small bond dimensions, yields substantial accelerations by almost two orders of magnitude, allowing 
calculations for CAS-sizes far beyond the current computational limits already feasible on single-node GPU accelerators. 
We expect future advances enabling DMRG calculations based on CAS spaces well beyond CAS(100,100) will soon be possible on multi-node, multi-GPU hardware architectures.
\\


\emph{Conclusions and outlook} 

In this work we report state-of-the-art performance results obtained on a single node NVIDIA DGX-H100 architecture via the spin adapted \textit{ab initio} density matrix renormalization group method. 
We observe a 2.5$\times$ speedup compared to a DGX-A100 node or equivalently an 80$\times$ speedup compared to an OpenMP parallelized 128 core CPU implementation.
These performance improvements reduce run times of typical DMRG calculations for quantum chemistry applications from many days to a few hours, making it possible to potentially apply DMRG  
routinely in scientific and industrial applications. We expect that with the development of even more advanced classical hardware in the near future, and their extension to shared-memory, multi-node multi-GPU architectures, tensor network calculations well beyond CAS(100,100) to be
achievable within hours. Such large CAS calculations may help elucidate the electronic mechanisms behind some of the most elusive chemical systems, such as multi-reference transition metal systems, catalysts, or metalloenzymes. 
We want to emphasise that a truly quantitatively correct description of such challenging problems requires a careful selection of the orbital active space and a balanced treatment of static and dynamic correlation effects. Chemists today largely rely on their intuition to find appropriate active spaces, and the question of finding the right 
one for a given system is a currently unsolved problem \cite{legeza_optimizing, Faulstich-2019b, autocas}. The ability to quickly iterate on different choices of large active spaces enables a more systematic search for an 
appropriate active space description. Combined with the ability to  
perform CASSCF calculations on larger active spaces in similarly short times, and robust approaches for CAS selection \cite{legeza_optimizing, Faulstich-2019b, autocas}, represents a significant step forward towards solving the CAS selection problem and obtaining quantitatively and qualitatively unambiguous results for strongly correlated systems.

Tensor network algorithms like DMRG \cite{white_density-matrix_1993}, projected entangled pair states (PEPS) \cite{verstraete2004renormalizationalgorithmsquantummanybody}, or the multi-scale entanglement renormalisation ansatz (MERA) \cite{vidal_mera} occupy a space at the intersection of classical and quantum computing, and are considered to be among the most powerful classical methods to treat strongly correlated and weakly entangled quantum systems. They play a key role in the quest for achieving quantum advantage, both for providing the best known classical answers to reference for many challenging problems \cite{Goings-2022,bellonzi2024feasibilityacceleratinghomogeneouscatalyst,zhou_what,chen_quantum,stoudenmire2023groversalgorithmoffersquantum, pan_solving} and as fundamental tools for building and testing quantum algorithms \cite{rudolph2022decompositionmatrixproductstates,Rudolph_2023}, simulating and understanding the real-time behavior of quantum hardware  \cite{Vidal-2003a, Vidal-2003b, Daley_2004,Haegeman_unifying}, and performing error correction \cite{ferris_tensor}. GPU accelerated tensor network algorithms can be expected to have significant impact in these areas in the years to come, and we expect our results to further boost community efforts aimed at the standardization and adoption of large-scale, GPU accelerated tensor contraction methods and libraries ~\cite{cecam-2024}.

\emph{Acknowledgments:}
This work has been supported by the Hungarian National Research, Development and Innovation Office (NKFIH) through Grant Nos.~K134983 and TKP2021-NVA-04,
by the Quantum Information National Laboratory
of Hungary and by
the Center for Scalable and Predictive methods
for Excitation and Correlated phenomena (SPEC),
funded as part of the Computational Chemical Sciences, 
FWP 70942,
by the U.S. Department of Energy (DOE), Office of Science, Office of Basic Energy Sciences, Division of Chemical Sciences, Geosciences, and Biosciences at Pacific Northwest National Laboratory.
\"O.L. acknowledges financial support
by the Hans Fischer Senior Fellowship programme funded by the Technical University
of Munich – Institute for Advanced Study.
The simulations were performed using Google Cloud Service and 
the Wigner Scientific Computational Laboratory (WSCLAB).


%
\end{document}